# Stacking-Dependent Band Gap and Quantum Transport in Trilayer Graphene

W. Bao[1], L. Jing[1], J. Velasco Jr.[1], Y. Lee[1], G. Liu[1,†], D. Tran[1], B. Standley[2], M. Aykol[3], S. B. Cronin[3], D. Smirnov[4], M. Koshino[5], E. McCann[6], M. Bockrath[1,2], and C.N. Lau[1,*]

[1] Department of Physics and Astronomy, University of California, Riverside, CA 92521

[2] Department of Applied Physics, California Institute of Technology, Pasadena, CA 91125

[3] Department of Electrical Engineering, University of Southern California, Los Angeles, CA 90089

[4] National High Magnetic Field Laboratory, Tallahassee, FL 32310

[5] Department of Physics, Tohoku University, Sendai, 980-8578, Japan

[6] Department of Physics, Lancaster University, Lancaster, LA1 4YB, United Kingdom

[†] Current address: Department of Chemistry and Biochemistry, University of California, Los Angeles, CA 90095.

*Email: lau@physics.ucr.edu

Graphene[1-3] is an extraordinary two-dimensional (2D) system with chiral charge carriers and fascinating electronic, mechanical and thermal properties[4,5]. In multilayer graphene[6,7], stacking order provides an important yet rarely-explored degree of freedom for tuning its electronic properties[8]. For instance, Bernal-stacked trilayer graphene (B-TLG) is semi-metallic with a tunable band overlap, and rhombohedral-stacked (r-TLG) is predicted to be semiconducting with a tunable band gap[9-17]. These multilayer graphenes are also expected to exhibit rich novel phenomena at low charge densities due to enhanced electronic interactions and competing symmetries. Here we demonstrate the dramatically different transport properties in TLG with different stacking orders, and the unexpected **spontaneous gap opening in charge neutral r-TLG. At the Dirac point, B-TLG remains metallic while r-TLG becomes insulating with an intrinsic interaction-driven gap ~6 meV. In magnetic fields, well-developed quantum Hall (QH) plateaus in r-TLG split into 3 branches at higher fields. Such splitting is a signature of Lifshitz transition, a topological change in the Fermi surface, that is found only in r-TLG. Our results underscore the rich interaction-induced phenomena in trilayer graphene with different stacking orders, and its potential towards electronic applications.**

TLG has two natural stable allotropes: (1) ABA or Bernal stacking, where the atoms of the topmost layer lie exactly on top of those of the bottom layer; and (2) ABC or rhombohedral stacking, where one sublattice of the top layer lies above the center of the hexagons in the bottom layer (Fig. 1a-b insets). This subtle distinction in stacking order results in a dramatic difference in band structure. The dispersion of B-TLG is a combination of the linear dispersion of single layer graphene (SLG) and the quadratic relation of bilayer (BLG) (Fig. 1a), whereas the dispersion of r-TLG is approximately cubic, with its conductance and valence bands touching at a point close to the highly symmetric K and K' points (Fig. 1b)[9,10,12,13]. These distinctive band structures are expected to give rise to different transport properties. For instance, owing to the cubic dispersion relation, r-TLG is expected to host stronger electronic interactions than B-BLG. This is because the interaction strength $r_s$ is approximately the ratio of the inter-electron Coulomb energy to the Fermi energy. In graphene, $r_s \propto n^{-(p-1)/2}$, where $n$ is charge density and $p$ is the power of the dispersion relation; $p$=1, 2, 3 for SLG, BLG and r-TLG, respectively[5]. Consequently, at low $n$, the interaction strength in r-TLG is significantly higher than that in SLG, BLG and B-TLG (the last can be considered as a combination of SLG and BLG)[18]. Hence, r-TLG potentially allows the observation of interaction-driven phenomena, *e.g.* spontaneous gap formation, that are not easily accessible in BLG or B-TLG. Thus we seek to experimentally explore the transport properties of TLG with different stacking orders.

Fig. 1c displays the two-terminal conductance $G$ of two suspended TLG devices with different stacking orders as a function of back gate voltage $V_g$ at $T$=1.5K. Both curves are "V"-shaped, characteristic of high mobility samples. Surprisingly, the two devices display drastically different minimum conductance $G_{min}$ at the charge neutrality point (CNP) – $G_{min}$ for B-TLG is ~50 µS, but <~1 µS for r-TLG. The strikingly large difference in minimum conductivity $\sigma_{min}$, as well as the very low $\sigma_{min}$ in certain high mobility samples, is unexpected and unique to TLG.

To systematically examine the effect of stacking order on $\sigma_{min}$, we investigated 21 substrate-supported and 22 suspended devices. After electrical measurements, the stacking order of the devices are identified using Raman spectroscopy[19]. In particular, the 2D peak of r-TLG is more asymmetric with a pronounced shoulder than that of B-TLG (Fig. 2a). Our findings are summarized in Fig. 2b, which plots $\sigma_{min}$ at $T$=4K vs. the field effect mobility $\mu$, revealing several interesting observations. For instance, for all B-TLG devices, $\sigma_{min}$ decreases with increasing sample mobility but remains finite, presumably because the same scattering mechanisms that yield low mobility also give rise to electron and hole puddles[20], hence smearing Dirac points and leading to higher $\sigma_{min}$. Amazingly, $\sigma_{min}$ for r-TLG devices is significantly smaller than B-TLG. The difference is at least a factor of 2 or 3 for substrate-supported devices, and becomes dramatic for suspended devices – $\sigma_{min, B-TLG}$ remains almost constant at ~100 µS for $\mu$>5x10$^4$ cm$^2$/Vs, while $\sigma_{min, r-TLG}$ ~ 0, suggesting the presence of metallic and insulating states, respectively.

The insulating state in neutral r-TLG is not anticipated from non-interacting electron pictures. To elucidate its nature and compare transport in TLG with different stacking orders, we investigate the devices' temperature dependence. Fig. 3a-b plots the $G(V_g)$ curve for B- and r-TLG devices, respectively, at $T$ between 1K and 120K. In both data sets, $G$ at small $n$ declines quickly with temperature, but stays almost constant or increases modestly for high $n$. The opposite $G(T)$ dependence in these two density regimes is similar to that observed in SLG[21,22], where the weak $T$-dependence at large $n$ is attributed to electron-phonon interaction[22].

At the CNP, $G_{min, B-TLG}$ displays a moderate $T$-dependence, typically decreasing by a factor of 2 -8 when $T$ is reduced from 200 to 1.4K (Fig. 3c). Data from another device with a small $T$-dependence is shown in Supplementary Information. Variable range hopping, which has an stretched exponential $T$-dependence, cannot adequately describe the data. We thus compare data in Fig. 3c to a model of thermally activated transport

$$G_{min} = G_0 + A e^{-E_A/k_B T} \qquad (1)$$

where $E_A$ is the activation energy, $k_B$ the Boltzmann constant, and $G_0$ and $A$ are fitting parameters. An adequate fit to Eq. (1) can be obtained by using $E_A$=25K, though the fit is not entirely satisfactory.

In contrast, $G_{min}$ of r-TLG displays an exceedingly strong temperature dependence – it decreases exponentially with $1/T$ by 2-3 orders of magnitude for 5<T<105K, crossing over to a constant value at lower temperatures (Fig. 3d). Using $E_A$=32.0 K, we obtain excellent agreement between the experimental data and Eq. (1), demonstrating that transport in r-TLG at the CNP occurs via thermal activation through an energy gap of $2E_A$ ~5.5 meV. The constant $G_0$ is sample-dependent, and decreases from 10 to 0.1 ∝S with improved mobility, indicating that it arises from scattering from residual impurities on the suspended membranes.

To gain further insight into the insulating state of r-TLG, we measure its differential conductance $dI/dV$ at $T$=300 mK *vs.* $V_g$ and source-drain bias $V$ (Fig. 3e). The resulting stability diagram reveals a diamond-like structure centred at CNP, where $dI/dV$ ~0 at $V$=0. For $V$>0.7 mV, *dI/dV* increases almost linearly with bias up to 15 mV with a width of ~7.5 mV, consistent with that determined from the activation energy. As *V* increases further to ~21 mV, *dI/dV* rises sharply to ~ 400 ∝S within 2mV (Fig. 3f). Such an abrupt jump in *dI/dV* strongly resembles the density of states for gapped phases such as superconductors or charge density waves, suggesting the presence of an intrinsic insulating state at the CNP with spontaneous symmetry breaking.

To sum our experimental findings: at *B*=0, we find that B-TLG remains metallic at the CNP, while r-TLG becomes insulating at low temperatures. $G_{min}$ of the latter is thermally activated for *T*>5K, with a gap-like feature in its *dI/dV* curve. Taken together, these results strongly suggest the presence of an intrinsic band gap in r-TLG. Such a gap is not anticipated from tight-binding calculations, and likely arises from electronic interactions, as expected from r-TLG's large interaction parameter $r_s$. For instance, a band gap may occur if spatial inversion symmetry is broken by strain or an external electric field, or if electronic interactions cause spontaneous symmetry breaking such as those predicted[23-25] or reported[26] for BLG.

Lastly, we focus on the transport characteristics of TLG devices in the quantum Hall (QH) regime. From tight-binding calculations that include only nearest-layer coupling, the Landau level (LL) spectrum for B-TLG is a superposition of those for SLG and BLG[9,27-29]:

$$E_{1,N}^{ABA} = \pm\sqrt{2\hbar v_F^2 eB|N|} \quad \text{and} \quad E_{2,N}^{ABA} = \pm\frac{\hbar v_F^2 eB}{t_\perp}\sqrt{N(N-1)} \qquad (2)$$

For r-TLG, the LL energies are given by[9,30]

$$E_N^{ABC} = \pm\frac{(2\hbar v_F^2 eB)^{3/2}}{t_\perp^2}\sqrt{N(N-1)(N-2)} \qquad (3)$$

In these expressions, *N* is an integer denoting the LL index, $v_F$~$10^6$ m/s is the Fermi velocity, *e* is the electron charge, *h* is Planck's constant and $t_\perp$~0.2-0.4 eV is the nearest-layer coupling energy. For both types of stacking order, the LL at zero energy is 12-fold degenerate, giving rise to quantized conductance plateaus with integer values ...-10, -6, 6, 10, 14... of $e^2/h$. When other interlayer and intralayer hopping terms are included, certain degeneracies could be broken[11,31], though the LL are expected to retain 4-fold degeneracy for B-TLG and 2-fold degeneracy for r-TLG.

Experimentally, in contrast to theoretical predictions above, all r-TLG and most B-TLG devices develop an insulating state at the Dirac point in finite $B$. This insulating state, with filling factor $\nu$=0, is often the first QH plateau that is energetically resolved. Fig. 4a plots the resistance at the Dirac point $R_{max}$ as a function of magnetic field $B$ for an r-TLG device at different temperatures. $R_{max}$ increases exponentially (with some localization-induced fluctuations) with increasing $B$ and decreasing $T$, spanning more than 3 orders of magnitude and reaching ~$10^9 \Omega$, which is the *de facto* limit of our measurement circuit.

For QH states at finite energies, Shubnikov-de Haas (SdH) oscillations start to emerge at $B$ as small as 0.2T. QH plateaus with filling factors $0 \leq \nu \leq 4$ can be identified[32]. Fig. 4b plots the plots the differentiated conductance $dG/dB$ $(B, V_g)$ of an r-TLG device, which allows the oscillations to be clearly discerned. The QH states appear as features radiating outward from the CNP at $B$=0. Since the device has a large aspect ratio (length/width ~3), we use conductance peaks to identify the filling factors of the QH plateaus[33], which are determined from their slopes in the $V_g$-$B$ plane: $\nu=nh/Be=\alpha V_g h/Be$, where $\alpha$, the gate coupling efficiency, is estimated to be $2.5 \times 10^{10}$ cm$^{-2}$V$^{-1}$ from geometrical consideration as well as the periods of SdH oscillations[32]. Using this relation, the features in Fig. 4b are determined to correspond to $\nu$=-30±1.2, -18±1,-9.3±0.5, 0, 9±0.5, 18±1, 30±1.2 and 42±2, respectively, as indicated on the figure. Fig. 4c plots the device conductance $G$ in units of $e^2/h$ as functions of $V_g$ taken at different values of $B$; when plotted against $\nu$, the 7 curves almost collapse into one, with plateaus at ~±9 and -18, respectively (Fig. 4d). We note that the conductance values approximately agree with the filling factors.

The emergence of filling factors at 9 and 18 are unexpected from Eq. (3). A close examination of Fig. 4b reveals yet another surprising feature -- some plateaus that appear at low fields unexpectedly disappear at higher values of $B$. For instance, the $\nu$=-18 feature is visible at $B$=0.25T and develops into a well-quantized plateau for 0.5<$B$<0.7T, yet it disappears for $B$>0.8T. Similarly, the $\nu$=-9 state is a well-developed plateau at 0.5T, but vanishes for $B$>1.5T. Instead, each of the $\nu$~±9, ±18 and -30 QH features splits into 3 branches at $V_g$~ 13-16V and $B$~ 0.6 – 1.3T. The splittings at $\nu$~9, -18 are indicated by the dotted circles in Fig. 4b. Such apparent 3-fold degeneracy of QH plateaus is highly surprising, and has not been observed in BLG or B-TLG devices with comparable mobility.

Such splittings are signatures of the Lifshitz transition (LT), a topological change in the Fermi surface as a function of electron doping or other parameters such as strain. For multilayer graphene, it may be induced by trigonal warping[28,30,34]: at very low $n$, the Fermi surface in r-TLG breaks up into 3-legged pockets, thus leading to triply degenerate LLs[30]; these LLs should split in higher $B$ or $n$, corresponding to the merging of the pockets at the LT. Indeed, the observed splittings occur at $|V_g|$~15V and $B$~1T, within 60% of the theoretically predicted values. The overall device behavior is in semi-quantitative agreement with theoretical simulations of r-TLG's density of states (Fig. 4e), which is satisfactory, considering that the simulation uses bulk graphite parameter values that are likely different for sheets of atomic thickness. We note that the biggest discrepancy between the data and simulation lies in the filling factor of the first non-zero plateau. Theoretically, one expects the $\nu$=6 plateau to be the most energetically stable; however, $\nu$~9 was observed instead, suggesting the presence of large valley and spin splitting.

In conclusion, our comprehensive study of Bernal and rhombohedrally-stacked trilayer devices reveal a number of fascinating phenomena, including spontaneous gap opening in undoped r-TLG, an insulating $\nu=0$ QH state, and Lifshitz transition induced by trigonal warping. This opens the door to explore a number of interesting questions, such as the nature of the insulating state in r-TLG , the unexpected $\nu=0$ and $\nu=9$ QH states, and transport across Bernal-rhombohedral stacking domains, and could enable new graphene electronics based on band gap and stacking-order engineering.

**Materials and Methods**

Graphene devices are fabricated by shadow mask evaporation of electrodes onto graphene sheets that are either supported on substrates or suspended across pre-defined trenches in $Si/SiO_2$ substrates[35]. These devices have no contaminants introduced by lithographical processes, with field effect mobility $\propto$ ranging from 210 to 1900 $cm^2$/Vs for non-suspended devices, and 5000 to 280,000 for suspended samples, which are significantly higher than those fabricated by lithography. We measure their electrical properties using standard lock-in techniques in a $He^3$ or pumped $He^4$ cryostat.

**References**

1 Zhang, Y. B., Tan, Y. W., Stormer, H. L. & Kim, P. Experimental observation of the quantum Hall effect and Berry's phase in graphene. *Nature* **438**, 201-204 (2005).

2 Novoselov, K. S. *et al.* Two-dimensional gas of massless Dirac fermions in graphene. *Nature* **438**, 197-200 (2005).

3 Novoselov, K. S. *et al.* Electric field effect in atomically thin carbon films. *Science* **306**, 666-669 (2004).

4 Castro Neto, A. H., Guinea, F., Peres, N. M. R., Novoselov, K. S. & Geim, A. K. The electronic properties of graphene. *Rev. Mod. Phys.* **81**, 109-162 (2009).

5 Das Sarma, S., Adam, S., Hwang, E. H. & Rossi, E. Electronic transport in two dimensional graphene. *Rev. Mod. Phys.*, to appear. arXiv:1003.4731v1002 (2010).

6 Liu, Y. P., Goolaup, S., Murapaka, C., Lew, W. S. & Wong, S. K. Effect of Magnetic Field on the Electronic Transport in Trilayer Graphene. *ACS Nano* **4**, 7087-7092 (2010).

7 Zhu, W. J., Perebeinos, V., Freitag, M. & Avouris, P. Carrier scattering, mobilities, and electrostatic potential in monolayer, bilayer, and trilayer graphene. *Phys. Rev. B* **80** (2009).

8 Mak, K. F., Shan, J. & Heinz, T. F. Electronic structure of few-layer graphene: experimental demonstration of strong dependence on stacking sequence. *Phys. Rev. Lett.* **104**, 176404 (2010).

9 Guinea, F., Castro Neto, A. H. & Peres, N. M. R. Electronic states and Landau levels in graphene stacks. *Phys. Rev. B* **73**, 245426 (2006).

10 Aoki, M. & Amawashi, H. Dependence of band structures on stacking and field in layered graphene. *Sol. State Commun.* **142**, 123-127 (2007).

11  Koshino, M. & McCann, E. Gate-induced interlayer asymmetry in ABA-stacked trilayer graphene. *Phys. Rev. B* **79**, 125443 (2009).

12  Craciun, M. F. *et al.* Trilayer graphene is a semimetal with a gate-tunable band overlap. *Nature Nanotechnol.* **4**, 383-388 (2009).

13  Zhang, F., Sahu, B., Min, H. K. & MacDonald, A. H. Band structure of ABC-stacked graphene trilayers. *Phys. Rev. B* **82**, 035409 (2010).

14  Manes, J. L., Guinea, F. & Vozmediano, M. A. H. Existence and topological stability of Fermi points in multilayered graphene. *Phys. Rev. B* **75**, 155424 (2007).

15  Partoens, B. & Peeters, F. M. From graphene to graphite: Electronic structure around the K point. *Phys. Rev. B* **74**, 075404 (2006).

16  Latil, S. & Henrard, L. Charge carriers in few-layer graphene films. *Phys. Rev. Lett.* **97**, 036803 (2006).

17  Koshino, M. Interlayer screening effect in graphene multilayers with ABA and ABC stacking. *Phys. Rev. B* **81**, 125304 (2010).

18  Zhang, F., Jung, J., Fiete, G. A., Niu, Q. A. & MacDonald, A. H. Spontaneous quantum Hall states in chirally stacked few-layer graphene systems. *Phys. Rev. Lett.* **106**, 156801 (2011).

19  Lui, C. H. *et al.* Imaging stacking order in few-layer graphene. *Nano Lett.* **11**, 164-169 (2011).

20  Martin, J. *et al.* Observation of electron-hole puddles in graphene using a scanning single electron transistor. *Nat. Phys.* **4**, 144-148 (2008).

21  Du, X., Skachko, I., Barker, A. & Andrei, E. Y. Approaching ballistic transport in suspended graphene. *Nat. Nanotechnol.* **3**, 491-495 (2008).

22   Bolotin, K. I., Sikes, K. J., Hone, J., Stormer, H. L. & Kim, P. Temperature-dependent transport in suspended graphene. *Phys. Rev. Lett.* **101**, 096802 (2008).

23   Nandkishore, R. & Levitov, L. Flavor symmetry and competing orders in bilayer graphene. *preprint*, arXiv:1002.1966v1001 (2010).

24   Vafek, O. & Yang, K. Many-body instability of Coulomb interacting bilayer graphene: Renormalization group approach. *Phys. Rev. B* **81**, 041401 (2010).

25   Zhang, F., Min, H., Polini, M. & MacDonald, A. H. Spontaneous inversion symmetry breaking in graphene bilayers. *Phys. Rev. B* **81**, 041402 (2010).

26   Weitz, R. T., Allen, M. T., Feldman, B. E., Martin, J. & Yacoby, A. Broken-symmetry states in doubly gated suspended bilayer graphene. *Science* **330**, 812-816 (2010).

27   McClure, J. W. Diamagnetism of Graphite. *Phys. Rev.* **104**, 666-671 (1956).

28   McCann, E. & Fal'ko, V. I. Landau-level degeneracy and quantum hall effect in a graphite bilayer. *Phys. Rev. Lett.* **96**, 086805 (2006).

29   Ezawa, M. Intrinsic Zeeman effect in graphene. *J. Phys. Soc. Jpn.* **76**, 094701 (2007).

30   Koshino, M. & McCann, E. Trigonal warping and Berry's phase N pi in ABC-stacked multilayer graphene. *Phys. Rev. B* **80**, 165409 (2009).

31   McCann, E. & Koshino, M. Spin-orbit coupling and broken spin degeneracy in multilayer graphene. *Phys. Rev. B* **81**, 241409 (2010).

32   Bao, W. *et al.* Magnetoconductance oscillations and evidence for fractional quantum Hall states in suspended bilayer and trilayer graphene *Phys. Rev. Lett.* **105**, 246601 (2010).

33   Abanin, D. A. & Levitov, L. S. Conformal invariance and shape-dependent conductance of graphene samples. *Phys. Rev. B* **78**, 035416 (2008).

34  Lemonik, Y., Aleiner, I. L., Toke, C. & Fal'ko, V. I. Spontaneous symmetry breaking and Lifshitz transition in bilayer graphene. *Phys. Rev. B* **82**, 201408 (2010).

35  Bao, W. Z. *et al.* Lithography-free fabrication of high quality substrate-supported and freestanding graphene devices. *Nano Research* **3**, 98-102 (2010).

**Supplementary Information** is linked to the online version of the paper at http://www.nature.com/naturephysics.

**Author Contributions**

C.N.L and W.B. conceived the experiments; W.B. and D. T. isolated and identified graphene sheets; W.B., L.J., Y.J., J.V., G.L, B.S. and D.S. performed transport measurements; W.B., L.J., M.A. and S.C. performed Raman measurements; C.N.L, M.B., W.B., L.J. and J.V. interpreted and analyzed the data; M.K. and E.M. interpreted data and performed theoretical calculations; C.N.L., M.B., W.B. and E.M. co-wrote the paper. All authors discussed the results and commented on the manuscript.

**Acknowledgements** This work was supported in part by ONR/DMEA H94003-10-2-1003, NSF CAREER DMR/0748910, ONR N00014-09-1-0724, and the FENA Focus Center. D.S. acknowledges the support by NHMFL UCGP #5068. The trenches are fabricated at UCSB. Part of this work was performed at NHMFL that is supported by NSF/DMR-0654118, the State of Florida, and DOE. S.C. acknowledges the support by ONR/N00014-10-1-0511. M.K and E. M. acknowledge the support by JST-EPSRC EP/H025804/1.

**Figure Captions**

**Figure 1 Characteristics of suspended B-TLG (ABA) and r-TLG (ABC) devices and their band structures. a-b,** Band structures (main panel) and schematics (inset) of B- and r-stacked TLG, respectively. **c**, $G(V_g)$ for two different suspended TLG devices at T=1.5K. Upper inset: $R(V_g)$ in log-linear scales for the same devices. Lower inset: SEM image of a suspended graphene device. Scale Bar: 2 $\mu$m.

**Figure 2. Different Raman and transport characteristics of B- and r-TLG. a**, Raman spectroscopy of TLG with different stacking orders. **b**, Minimum conductivity $\sigma_{min}$ vs. field effect mobility $\mu$ at 4K for suspended and non-suspended graphene devices.

**Figure 3. Transport data from B- and r-TLG devices. a-b**, $G(V_g)$ for B- and r-TLG devices, respectively, taken at different temperatures. Inset in **b**: Zoom-in plot of $G(V_g)$ curves at $T$=0.6, 0.8, 5.2, 7.7 and 10K (bottom to top). The curves at 0.6 and 0.8 K are indistinguishable. **c-d**, $G_{min}$ vs. $1/T$. The blue lines are best fits to Eq. (1), with $E_A$=25K and 32K for B-stacked and r-stacked TLG, respectively. Insets: $G_{min}(T)$ for the same data sets shown in the main panels. **e**, $dI/dV(V_g, V)$ for an r-TLG at $B$=0 and $T$=300mK. **f**. Line trace of **e** at $V_g$=0.

**Figure 4. Magnetotransport data of a r-TLG device. a**, $R_{max}(B)$ at different temperatures **b**, $dG/dB(V_g, B)$. The numbers indicate the filling factors of the features. **c-d**, $G(V_g)$ and $G(\nu)$ at $T$=1.5 K and $B$=0.5(blue), 0.6(cyan), 0.8(green), 1(yellow), 1.2(orange), 1.5(red), and 1.7 T(magenta). 1.25 k$\Omega$ has been subtracted from device resistance to account for the contact

resistance and line resistance of the cryostat. **e**, Calculated density of states for r-TLG *vs.* $B$ and $n$.

**Figure 1. Characteristics of suspended B-TLG (ABA) and r-TLG (ABC) devices and their band structures. a-b,** Band structures (main panel) and schematics (inset) of ABA- and ABC-stacked TLG, respectively. **c**, $G(V_g)$ for two different suspended TLG devices at T=1.5K. Upper inset: $R(V_g)$ in log-linear scales for the same devices. Lower inset: SEM image of a suspended graphene device. Scale Bar: 2 μm.

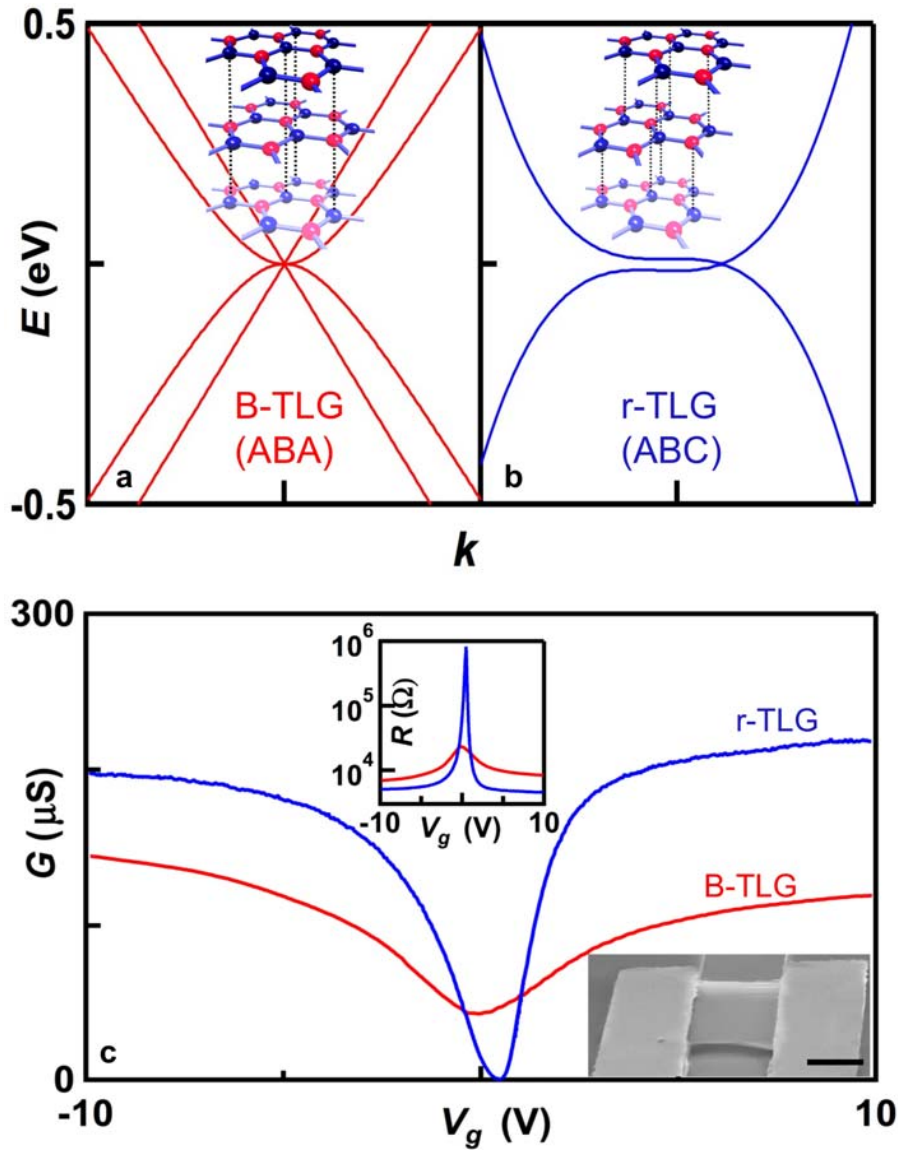

**Figure 2. Different Raman and transport characteristics of B- and r-TLG. a**, Raman spectroscopy of TLG with different stacking orders. **b**, Minimum conductivity $\sigma_{min}$ vs. field effect mobility $\propto$ at 4K for suspended and non-suspended graphene devices.

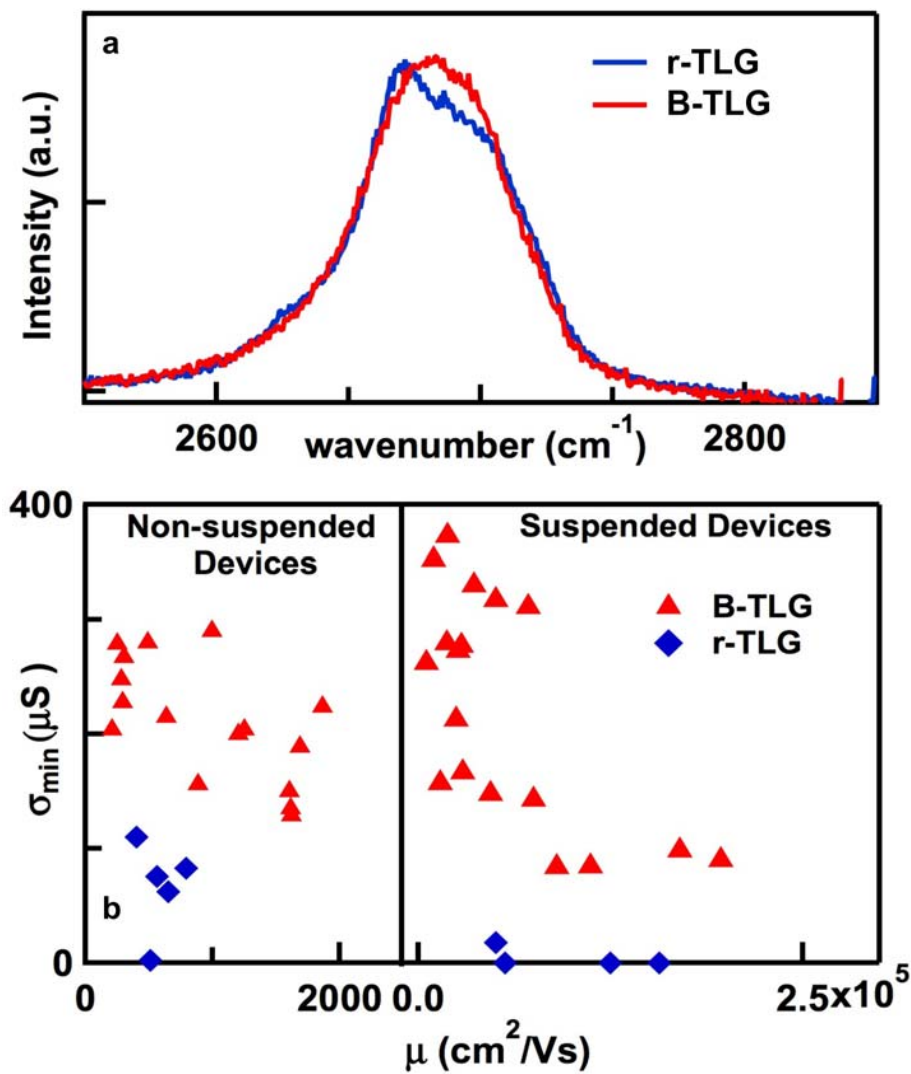

**Figure 3. Transport data from B- and r-TLG devices. a-b**, $G(V_g)$ for B- and r-TLG devices, respectively, taken at different temperatures. Inset in **b**: Zoom-in plot of $G(V_g)$ curves at $T$=0.6, 0.8, 5.2, 7.7 and 10K (bottom to top). The curves at 0.6 and 0.8 K are indistinguishable. **c-d**, $G_{min}$ vs. $1/T$. The blue lines are best fits to Eq. (1), with $E_A$=25K and 32K for B-stacked and r-stacked TLG, respectively. Insets: $G_{min}(T)$ for the same data sets shown in the main panels. **e**, $dI/dV(V_g,V)$ for an r-TLG at $B$=0 and $T$=300mK. **f**. Line trace of **e** at $V_g$=0.

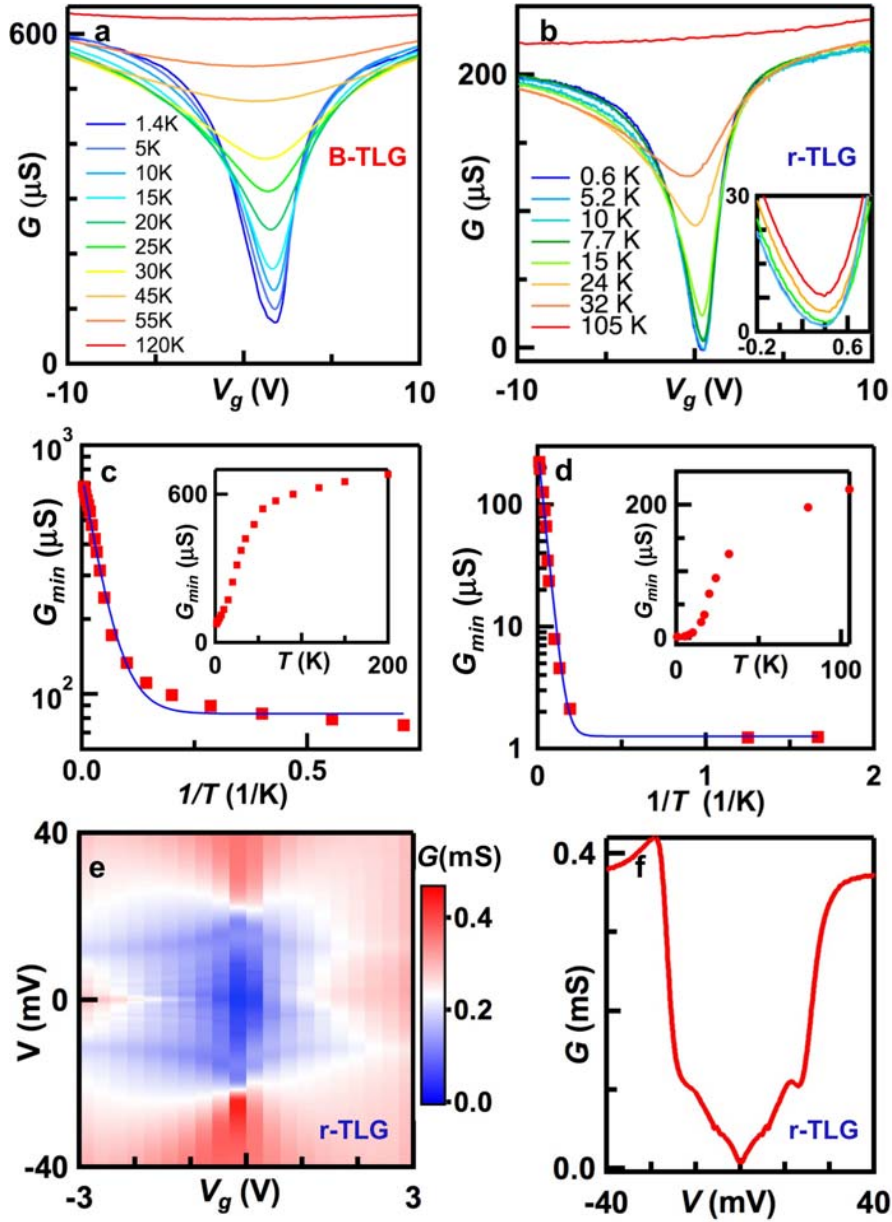

**Figure 4. Magnetotransport data of a r-TLG device. a**, $R_{max}(B)$ at different temperatures **b**, $dG/dB(V_g, B)$. The numbers indicate the filling factors of the features. **c-d**, $G(V_g)$ and $G(\nu)$ at $T$=1.5 K and $B$=0.5(blue), 0.6(cyan), 0.8(green), 1(yellow), 1.2(orange), 1.5(red), and 1.7 T(magenta). 1.25 kΩ has been subtracted from device resistance to account for the contact resistance and line resistance of the cryostat. **e**, Calculated density of states for r-TLG *vs.* $B$ and $n$.

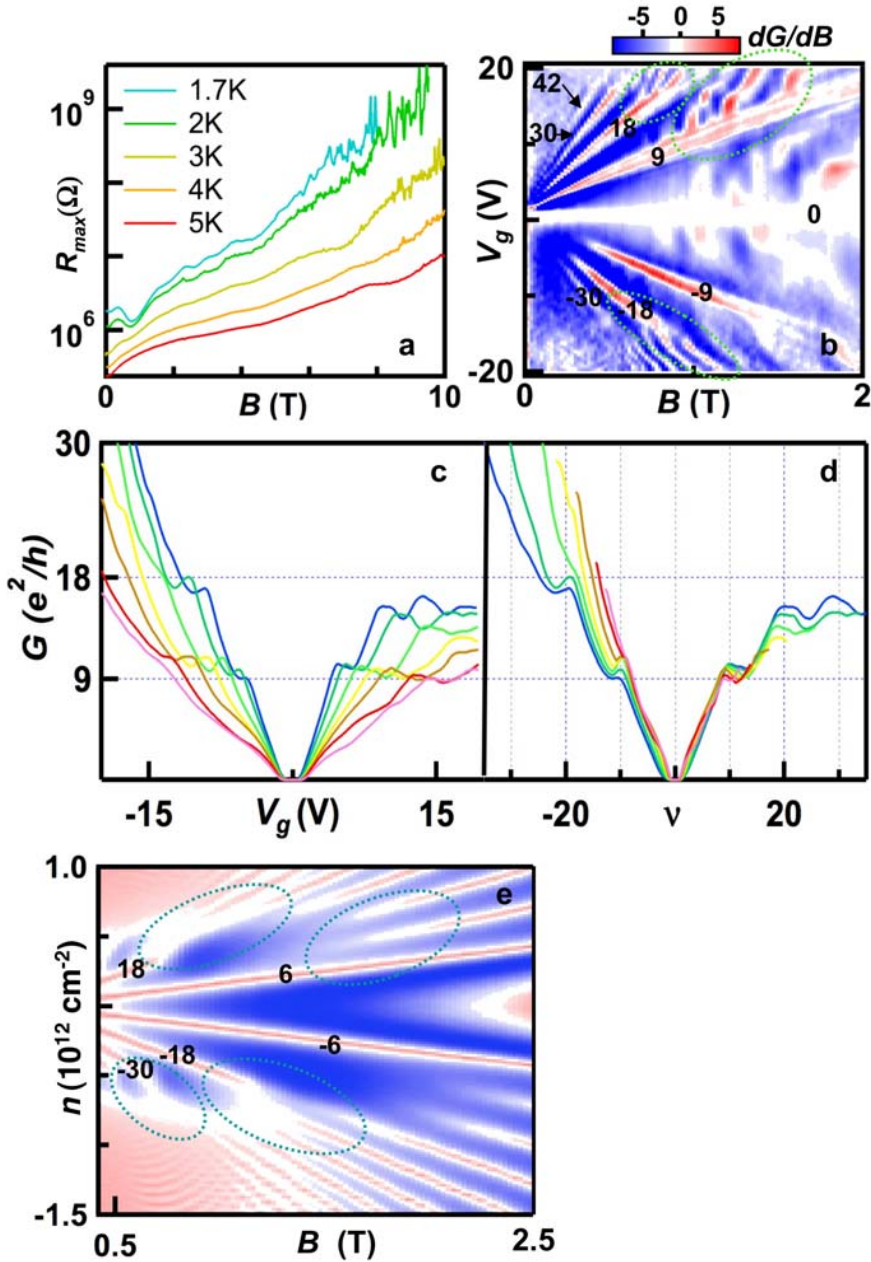